\documentclass[onecolumn]{article}

\usepackage[dvips]{graphicx}
\usepackage{amssymb,amsfonts,amsmath}

\date{}

\begin{document}

\title{Augmented Sparse Reconstruction of Protein
Signaling Networks.}

\author{\renewcommand{\thefootnote}{\alph{footnote}} D. Napoletani \thanks{Center for Applied Proteomics and Molecular Medicine, George Mason University, Manassas, VA
20110 USA.} \thanks{ Corresponding author present address:
Department of Mathematical Sciences, 4400 University Drive MS 3F2,
George Mason University, Fairfax, VA 22030 USA. email:
dnapolet@gmu.edu. Phone number: 703-993-4269, Fax number:
703-993-1491}, T. Sauer
\thanks{Department of Mathematical Sciences, George
Mason University, Fairfax, VA 22030 USA} , D. C. Struppa
\thanks{Department of Mathematics and Computer Science, Chapman
University, Orange, CA 92866 USA}, E. Petricoin$^a$, L. Liotta
$^a$}

\maketitle

\begin{abstract}
The problem of reconstructing and identifying intracellular
protein signaling and biochemical networks is of critical
importance in biology today. We sought to develop a mathematical
approach to this problem using, as a test case, one of the most
well-studied and clinically important signaling networks in
biology today, the epidermal growth factor receptor (EGFR) driven
signaling cascade. More specifically, we suggest a method,
augmented sparse reconstruction, for the identification of links
among nodes of ordinary differential equation (ODE) networks from
a small set of trajectories with different initial conditions. Our
method builds a system of representation by using a collection of
integrals of all given trajectories and by attenuating block of
terms in the representation itself. The system of representation
is then augmented with random vectors, and $l_1$ minimization is
used to find sparse representations for the dynamical interactions
of each node. Augmentation by random vectors is crucial, since
sparsity alone is not able to handle the large error-in-variables
in the representation. Augmented sparse reconstruction allows to
consider potentially very large spaces of models and it is able to
detect with high accuracy the few relevant links among nodes, even
when moderate noise is added to the measured trajectories. After
showing the performance of our method on a model of the EGFR
protein network, we sketch briefly the potential future
therapeutic applications of this approach.

\end{abstract}

\section{Introduction}

The problem of reconstructing a network of interacting variables
from a small set of data generated by the network itself has
attracted considerable attention especially since this problem
arises so naturally in genomics, proteomics and more generally
system biology problems (see for example \cite{Voit},
\cite{Voit2}, \cite{Husmeier}, \cite{Rogersetal},
\cite{Nachmanetal}, \cite{Collins1}). In particular, the ability
to reconstruct and identify intracellular protein signaling and
biochemical networks is of critical importance in modern biology.
However, the ability to dynamically measure and collect enough
data from every protein/node within the network is impossible with
current methodologies. We sought to develop a mathematical
approach to this problem using one of the most well-studied and
clinically important signaling networks in biology today, the
epidermal growth factor receptor (EGFR) driven signaling cascade
\cite{Araujoetal}.

Interestingly, it is widely believed, and proven in some cases,
that biological networks are scale free networks with a few
variables (hubs) very connected to many others and most variables
interacting only with a few others \cite{Albert}. Even the hubs do
not interact with more than a dozen other variables in most
reliable models, so that effectively we can say that these
networks are {\it sparse}, with respect to the total number of all
possible connections among variables. Such information  can
greatly help in reconstructing the network itself, as shown in
\cite{Collins1}, \cite{Collins2}, \cite{Collins3}.

Many current algorithms to reconstruct networks from expression
data are based on the application of powerful Bayesian methods
after the seminal work in \cite{Friedmanetal}, but, as noted in
\cite{Rogersetal} (see also \cite{Zaketal}), these methods do not
perform well with the limited amount of data that can be generated
by microarray technologies. This limitation is especially pertinent
for protein expression data. The other widely used approach for
network reconstruction is based on parameter estimation of
dynamical system models of the networks themselves \cite{Voit}. The
fundamental difficulty of such approach is the very large number of
parameters and reaction rates that need to be estimated
\cite{Voit2}, and this, again, leads to an inability to work
efficiently with the limited data generated by microarrays and time
series of expression profiles. Another viable alterative when
analyzing microarray data is to simply perform some type of
clustering analysis such as hierarchical or $K$-means clustering
\cite{Clustering}, or the recent exemplars clustering technique
\cite{Frey}. Clustering techniques do not require very large data
sets to be applied, but they only identify similarly activated
variables, and do not provide a causal understanding of the network
structure.

To address the need for specialized network reconstruction methods
that can work for the limited data generated by experiments, we
restrict our attention in this work to ordinary differential
equation (ODE) models of protein signaling networks of the form
$\dot x=f(x)$, where $x$ is the vector of variables in the system
and $\dot x$ its componentwise derivative. While assuming the
plausibility for biological networks of a dynamical system model
is a well established approach in the literature
(\cite{Voit},\cite{Voit2}), we believe that exact modeling and
parameter estimation for such models is not the most efficient way
to find how quantities interact in biological systems, partly
because parameter estimation is so difficult in noisy environments
and with small data sets to work with. In addition, biological
systems adapt their very network structure over time, especially
in the presence of diseases. It is likely more effective to search
for equivalent, indistinguishable, classes of models \cite{Smith}
that project to the same network structure, in the sense that they
give rise to trajectories that are qualitatively similar and they
have similar overall topology of the connections among nodes

With this general approach in mind, we ask in this paper the
following question: whether the structure of sparse ODE networks
can be inferred from a small set of trajectories with different
initial conditions generated by the system itself. We show that,
for a specific realistic case of ODE modeling of protein networks,
it is possible to expand and adapt ideas from the theory of sparse
regression (lasso) and signal reconstruction by $l_1$ minimization
(\cite{Tib}, \cite{CDS}, \cite{Tib2} chapter 3, \cite{Donoho1}),
to develop a method that reconstructs a significant portion of
these networks with good accuracy even in the presence of moderate
noise of intensity up to $20\%$ of the maximum values of the
trajectories. Our method builds a system of representation by
using a collection of integrals of all given trajectories and by
attenuating block of terms in the representation itself. The
system of representation is then augmented with random vectors,
and $l_1$ minimization is used to find sparse representations for
the dynamical interactions of each node. Augmentation by random
vectors is crucial in the context of network reconstruction, since
sparsity alone is not able to handle the large error-in-variables
in the representation when trajectories are very noisy.

One of the main strengths of our method is the ability to sharply
distinguish relevant links, so that the rate of false links that
are detected can be made very low. This is important in practice
since it is difficult and expensive to follow up and validate
experimentally potential links among proteins that are inferred by
computational means \cite{Huetal}.

The paper \cite{Collins2} is a significant antecedent to our work,
since in that paper the authors use a hybrid singular value
decomposition (SVD) and $l_1$ minimization to find a sparse linear
model that fits oligonucleotide microarray data. The $l_1$
minimization is used in that paper as a postprocessing of the
reverse-engineering performed by the SVD. A similar
preconditioning for large models is implemented in the recent
paper \cite{Tib3} and applied to several examples including
microarray data. In this paper we will show how $l_1$ minimization
methods can be modified to {\it directly} approach network
reconstruction and model identification problems, without any
preprocessing, for realistic, very limited sampling of the data
and significant noise levels, even when large spaces of non-linear
models are considered. Moreover the results for the EGFR network
show that our method can recover the topology of relatively small
protein networks. We do not require explicit estimation of the
noise level in the trajectories and we do not need multiple
trajectories with same initial conditions to estimate the true
trajectories in the presence of noise.

In section 2 we show how to apply $l_1$ minimization methods to
reconstruct sparse ODE networks, stressing the specific steps that
are necessary in the network setting. In section 3 we apply the
algorithm sketched in section 2 to one particular protein network,
the EGFR model as described in \cite{Araujoetal}.

We do not explore the biological significance of this network,
strongly related to cell proliferation, but we mention here that
it plays a significant role in cancer development \cite{EGFR}, so
it is considered an ideal target for fine tuned potential
therapies that do not impact the body at the systemic level. We
will briefly mention some possible directions of research related
to the medical applications of network mapping at the end of the
paper.

\section{Methods}

\subsection{Sparse Signal Processing}

Suppose we have a discrete function $F(n)$, $n=1,...,N$ and a
collection of functions $\mathcal G=\{g_1(n),...,g_M(n),\,\,
n=1,...,N\}$ with $M>>N$. Then in general the representation of
$F$ in terms of $\mathcal G$ will not be unique, meaning that
there will be many ways to write $F$ as $F(n)=\sum_{m=1}^M
a_mg_m(n)$, $n=1,...,N.$ An important question when trying to
extract the significant features of $F$ with respect to $\mathcal
G$ is to find, among the many possible representation of form as
in equation (1), the one that is the most `sparse', that is the
representation that has as many zero coefficients $a_m$ as
possible. This problem is in general very difficult, but we can
use linear programming techniques to find approximate sparse
representations, that is, representations that have just a few
large coefficients and many very small ones.

We briefly introduce this type of approximation to sparse
solutions here following mostly \cite{Mallat}, section 9.5.1 and
we refer to \cite{Tib}, \cite{CDS}, \cite{Tib2}, \cite{Donoho1}
and \cite{Donoho2} for a thorough analysis of the relations
between $l_1$ optimization and sparsity. The key idea is to
realize that if we minimize the 1-norm of the coefficients
$|a|=\sum_{m=1}^M |a_m|$, this implies that the total energy of
the coefficients is concentrated in just a few of them. We can
gain an intuition on this by noting that a minimization of the
1-norm reduces cancellations among different elements of $\mathcal
G$, since these cancellations increase the 1-norm.

Note that the problem $\min(\sum_{m=1}^M |a_m|),$ subject to
$F(n)=\sum_{m=1}^M a_mg_m(n)$, $n=1,...,N$, is equivalent to the
problem $\min(\sum_{p=1}^{2M} x_p)$, subject to $F(n)=\sum_{p=1}^M
x_mg_m(n)-\sum_{p=M+1}^{2M} x_mg_m(n)$, with $x_p>0$ for every
$p=1,...,2M$ and $x_p-x_{p+M}=a_p$. The linear optimization
problem defined by the last two equations can be easily put in the
standard format of linear programming problems, so that a solution
can be quickly obtained using one of several algorithms
\cite{Lustigetal}.

Given therefore a discrete signal of length $N$ and a collection
of $M$ signals $\mathcal G$ with $M>>N$ we can easily find
approximate sparse representations for $F$ in $\mathcal G$. This
result, first exposed in \cite{CDS}, was the inspiration of a
series of works that showed the great potential of $l_1$
minimization in signal processing, see for example the recent work
in \cite{CandesTao}, \cite{CandesRomTao2}. Regression using $l_1$
optimization has also been used extensively and to great effect in
statistical learning for model identification under the name of
lasso, after the pioneering work in \cite{Tib}, see \cite{Tib2}
chapter 3 for a up to date review of use of the technique.

We will see in the next subsection the crucial adjustments that
are required to make $l_1$ optimization effective and robust in
network reconstruction problems.

\subsection{Augmented Sparse Networks}

Let us now write explicitly the general form of the dynamical
systems of interest. Signaling networks arising from protein
interactions, even though often nonlinear, are often modeled with
differential equations that contain simple analytical forms that
contain power function terms of variables $x_1,...,x_N$ of the
type $x^{\alpha}=x_1^{\alpha_1}x_2^{\alpha_2}...x_n^{\alpha_n}$,
and hyperbolic terms of the type $\frac{x_i}{C+x_i}$, that take
into consideration the presence of slow enzymatic kinetics (see
\cite{Voit}, \cite{Araujoetal} and references therein). The
assumption of sparsity of links among the nodes implies that many
of the $\alpha_i$ are actually zero. In this paper we assume for
simplicity that the right hand side of the dynamical system that
we try to model has polynomial terms up to degree $d=2$, and
hyperbolic terms $\frac{x_i}{C+x_i}$, $C>0$, more specifically we
sample $C$ at uniform intervals of length $\bar c$ in a range
$[0,S\bar c]$ of interest with $S$ some large positive integer. If
we denote by $\dot{x}_i$ the time derivative of $x_i$, we consider
models of the form:
\begin{equation}
\dot{x}_n=a_0+\sum_{i=1}^N a_ix_i+\sum_{i=1}^N\sum_{j=1}^N
b_{ij}x_ix_j+\sum_{i=1}^N\sum_{s=1}^S \frac{x_i}{s\bar c+x_i}
\end{equation}
where $n=1,...,N$. We can in principle consider a model that
contains on the right had side all monomials $x_i^{\alpha_i}$, all
binomials $x_i^{\alpha_i}x_j^{\alpha_j}$, all the way to
$x_1^{\alpha_1}x_2^{\alpha_2}...x_n^{\alpha_n}$, where the
exponents have norm $|\alpha_i|$ less than a constant $A$ and we
assume a uniform sampling of the exponents. This would be a
general setting compatible with the modeling approach taken in
\cite{Voit}, however, the main  complication of such general
models is already severe in our quadratic model with hyperbolic
terms: when we have many nodes in the network, the combinatorial
explosion of terms makes parameter fitting very difficult in the
case only a limited amount of data is available on the dynamics of
each node.

A possible way to approach the fitting problem implicit in
equation (1) is to find the model that minimize the $l_1$ norm of
the parameters of the terms in the equation. From the background
material summarized in the previous subsection, we know that $l_1$
optimization leads to a sparse representation of signals with very
few terms with non-zero parameters, and that the optimization
itself can be performed with linear programming techniques
\cite{CDS} .

Since we noted in the introduction that actual biological networks
are generally sparse, the $l_1$ fitting method should, in
principle, improve our ability to find the actual links among
nodes. Exact parameter fitting is difficult in this case as well
and we will see in the results section that direct application of
the $l_1$ fitting as used in signal processing leads to very poor
results.

To be specific, we assume that we sample variables $x_1,...x_N$
and that we have {\it several} trajectories $x_{n,r}$ $r=1,...,R$
with $R$ different initial conditions. We denote by
$\dot{x}_1,...,\dot{x}_N$  the respective derivatives at each of
the sampled points. If we write $X_n=[x_{n,1},...,x_{n,R}]$,
$\dot{X}_n=[\dot{x}_{n,1},...,\dot{x}_{n,R}]$, and we denote by
$J$ the unit vector of same length as $X_i$, a formal substitution
in equation (1) of $x_n$ with $X_n$ and $\dot{x}_n$ with
$\dot{X}_n$ leads in effect to a problem of representation of
discrete signals $\dot{X}_n$ in terms of the collection of signals
$\mathcal X=\{J, X_i, X_jX_k, \frac{X_i}{s\bar c+X_i}\}$ with
$i,j,k=1,...,N$, $s=1,...,S$. This way of stating the problem of
reconstructing a specific network of the form (1) highlights the
potential of applying the $l_1$ sparsity techniques to recover the
effective system from a collection of different trajectories. Now
the first requirement for applying the $l_1$ method is to have an
underdetermined system with the cardinality $M$ of $\mathcal X$
such that $M>>L$, where we denote by $L$ the length of vectors in
$\mathcal X$. However a direct application of $l_1$ optimization
to the network data will not work in the presence of high noise
and for very limited data.  As much as sparsity is a powerful
device to explore signal representations, it is not able by itself
to deal with the large error-in-variables in the representation
generated by the system trajectories. There are some crucial
modifications that are necessary to get useful reconstruction
results on protein networks, we term them model augmentation,
attenuation of blocks of terms, and integral modeling.

{\bf Model Augmentation with Random Terms:} The main issue that
prevents an accurate reconstruction of the network is the presence
of noise in the trajectories. Because of such noise the
representation system needs to account for large
errors-in-variables \cite{Vossetal} when fitting the models on the
noisy data. To gain a better sense of this problem, denote the
noisy measurements of $X_i$ and $X_j$ as $\tilde X_i=X_i+N_i$ and
$\tilde X_j=X_j+N_j$ respectively, and assume that the
differential model includes a term $X_iX_j$ in the representation
of some $\dot{X}_n$. This means that when we represent $\dot{X}_n$
in $\mathcal X$ we do not just want a sparse representation, but
we would like the term $\tilde X_i \tilde X_j$ to appear in that
specific representation with large non zero coefficient. Now
$\tilde X_i \tilde X_j=X_iX_j+X_iN_j+X_jN_i+N_iN_j$, and we would
like the noisy residue $X_iN_j+X_jN_i+N_iN_j$ to `disappear', i.e.
to contribute marginally to the $l_1$ optimization. The way we
approached this problem is to go from the representation in
equation (1) to a representation:
\begin{eqnarray}
\dot{X}_n=a_0+\sum_{i=1}^N a_{i}X_i+\sum_{i=1}^N\sum_{j=1}^N
b_{ij}X_iX_j+
\\
\nonumber +\sum_{i=1}^N\sum_{s=1}^S \frac{X_i}{s\bar
c+X_i}+\sum_{g=1}^G n_g
\end{eqnarray}
where $n_g$ $g=1,..,G$ are discrete random vectors normally
distributed, scaled to have norm 1. We want $G$ much larger that
$L$ so that the energy of noisy residues like
$X_iN_j+X_jN_i+N_iN_j$ is uniformly distributed among all the
random vectors $n_g$, and the overall contribution to the $l_1$
norm of these noisy residues is small. Moreover a large value of
$G$ improves the conditioning of the corresponding linear
programming problem and therefore the speed of convergence to the
optimal solution. Note that $G$ is dependent on the particular
instance of problem that is given, and more specifically on the
type and number of trajectories and sample points in each
trajectory, but the performance of the method we describe in this
section is not strongly dependent on its specific value, as long
as $G>>L$. In Figure 5 we show numerical evidence of the
significance of adding random terms in the context of the
epidermal growth factor receptor case study.

This extension of the basic model has far reaching consequences,
since it assures that the new models are large enough to be able
to perform an approximate sparse minimization, strongly retaining
the dependence from the original terms of the `effective',
non-random model, while diffusing any potential noise in the data
among the random terms of equation (2). The non-random portion of
the matrix derived from the ODE network itself can be very ill
conditioned. In particular, the hyperbolic terms generated by the
same variable will be highly correlated among each other. There is
an intrinsic inability to fully control the representation matrix
generated by the trajectories and the error in variables that are
bound to appear when trajectories are very noisy. This is a
distinct characteristic of ODE reconstruction networks and one
that makes this work diverge in methodology and outlook from
standard $l_1$ signal reconstruction.

{\bf Attenuation of Block of terms:} We seek to have reconstructed
models with low complexity, that is, with terms of low degree, so
it is useful to enforce a way to explicitly suppress the terms
belonging to more complex blocks of terms such as quadratic and
hyperbolic ones. The large number of quadratic and hyperbolic
terms increases the chance, in a noisy setting, that several wrong
terms from these blocks are selected in the representation of each
node. By suppressing each of these blocks of terms we reduce the
chance of this wrong selection and we give more weight to linear
terms. Such suppression of higher complexity terms can be done by
using suitable attenuation coefficients. More specifically, we
choose to attenuate uniformly all terms in a block by a factor
$0<\beta<1$. Assuming that all vectors of the collection of terms
$\mathcal X$ were scaled to have $l_{2}$ norm equal to $1$, we
effectively multiply their inner product with any signal by
$\frac{1}{\beta}$, which is bigger than $1$, so the $l_1$
optimization will have the tendency to select fewer of them to
chose the representation with the minimal $l_1$ norm. This is
another interesting point specific to the modeling of networks.
Empirically, we find that this adjustment is important for
obtaining the very best results in the reconstruction of the
geometric structure of the network, for example an attenuation of
a factor of $\beta=0.5$ for both quadratic and hyperbolic terms
was near optimal for the epidermal growth factor receptor network.
The need of some attenuation is especially strong when we want
very few selected false links and the trajectories are very noisy.
We find that a wide range of small values of $\beta$ gives similar
reconstruction results, but the optimal selection of $\beta$ for
each different block, including a possible attenuation for the
block of random terms, is an open problem and we will explore
numerically this issue in a separate paper. Essentially, these
attenuation coefficients are one more device to keep the
errors-in-variables from generating false links in the computed
representation of each node, assuming that low degree and low
complexity terms are to be preferred.

{\bf Integral Modeling:} In a realistic reconstruction setting we
have few sample points and a relative noise that can be as high as
$20\%$ in the measured trajectories, making the estimation of the
derivatives very difficult. This problem transcends the specifics
of our approach and is a key issue in the study of experimentally
generated time series. Note that to use $l_1$ optimization, we
clearly do not need to use only local differential information. To
avoid the problem of direct estimation of derivatives in the
highly noisy cases, we note that the equations as (1) can be
written for $n=1,...,N$, in integral form as:
\begin{eqnarray}
x_n(t)-x_n(t_0)=a_0+\sum_{i=1}^N a_i \int_{t_0}^t x_i dt+
\\
\nonumber +\sum_{i=1}^N\sum_{j=1}^N b_{ij}\int_{t_0}^t x_ix_j dt+
\sum_{i=1}^N\sum_{s=1}^S \int_{t_0}^t \frac{x_i}{s\bar c+x_i}dt.
\end{eqnarray}
This integral representation avoids the implicit problem of
finding a good estimation of the derivative from a limited number
of samples of the trajectories. The relative noise in the
measurement of $x_n(t)-x_n(t_0)$ is comparable with the relative
noise of the time series $x_n$ itself when $t$ is far from $t_0$.
Moreover, for biological signals derived from proteomics and
genomics, variables often represent intensity or concentration
profiles that always assume positive values and therefore, in
these cases, we expect the integrals on the right hand side to be
dominated by the integrals of the true values of the variables,
when zero mean noise is added to them. A further advantage of
integral modeling is that we can easily estimate multiples of the
integrals on the right hand side of equation (3) by summing up the
samples that are given from $t_0$ to $t$, if sampling is uniform.
If sampling is not uniform, which is very often the case for
experimental data, we can scale the contribution of each summand
multiplying by the size of the corresponding sampling interval.

Note that the constant term $a_0$ was used simply as a term to
correct potential biases in (1), as it does not carry information
on the nodes' links, so we use it similarly in (3) and we do not
take its integral. The augmentation by random terms and the
attenuation of blocks of terms clearly can be applied to the
integral representation (3) as well. The system representation
that takes into account random augmentation, attenuation of blocks
of terms and integral modeling is the following:

\begin{eqnarray}
x_n(t)-x_n(t_0)=a_0+\sum_{i=1}^N a_i \int_{t_0}^t x_i dt+
\\
\nonumber +\beta_q\sum_{i=1}^N\sum_{j=1}^N b_{ij}\int_{t_0}^t
x_ix_j dt+\beta_h\sum_{i=1}^N\sum_{s=1}^S \int_{t_0}^t
\frac{x_i}{s\bar c+x_i}dt+\sum_{g=1}^G n_g.
\end{eqnarray}
where $\beta_q$ and $\beta_h$ are positive attenuation
coefficients for quadratic and hyperbolic terms, both smaller than
$1$. This representation is used in the actual network
reconstruction algorithm that we are going to describe now.

\subsection{Network Reconstruction Algorithm}

The observations in the previous subsection can be gathered into a
simple reconstruction algorithm based on $l_1$ optimization, which
we call augmented sparse reconstruction. We label the variables
involved in a slightly different way in the algorithm to highlight
the flexibility in the choice of the input for the algorithm.
Given trajectories from a sparse system that is believed to be of
a certain generic form, for each discretely sampled trajectory
$X_{n,r}$, $r=1,...,R$, let $\bar X_{n,r}$ be the vector
$X_{n,r}(t)-X_{n,r}(t_0)$ where $t$ takes all sampled values.
Moreover, for a given vector $g(t)$, $t=t_0,...,t_L$, let $I(g)$
be the vector whose $l$-th component is the sum $\sum_{i=0}^l
g(t_i)$, and let $J$ denote the unit vector. The basic process to
identify the nodes is the following:
\begin{itemize}

\item[{\bf A}] Suppose we are given $N$ node variables and that
for each variable it is possible to generate $R$ uniformly sampled
trajectories $X_{n,r}$ $r=1,...,R$ with different initial
conditions. Write $Y_n=[\bar X_{n,1},...,$$\bar X_{n,R}]$,
$G_n=[I(X_{n,1}),...,$ $I(X_{n,R})]$, $n=1,...,N$, $G_{ij}=[I(
X_{i,1}X_{j,1}),...,$$I(X_{i,R}X_{j,R})]$  and
$H_{js}=[I(\frac{X_{j,1}}{s\bar c+X_{j,1}}),...,$
$I(\frac{X_{j,R}}{s\bar c+X_{j,R}})]$, $s=1,2,...,S$ and $\bar c$
is the sampling interval for the hyperbolic terms. For each
$n=1,...,N$:

\item[{\bf B}] Choose an attenuation coefficient $\beta_q$ for the
quadratic terms and another one, $\beta_h$, for the hyperbolic
terms. Let $n_g$, $g=1,..,G$, be discrete random vectors normally
distributed scaled to have norm 1. Denote by $|\,|$ the $2$-norm
of a vector and let $\hat G_l$ be the matrix whose columns are all
the vectors $\frac{G_i}{|G_i|}$, $\hat G_{q}$ be the matrix whose
columns are all possible vectors $\frac{G_{ij}}{|G_{ij}|}$ and
$\hat H$ be the matrix whose columns are all allowed hyperbolic
terms $\frac{H_{is}}{|H_{is}|}$. Let $N_G$ be the matrix whose
columns are the random vectors $n_g$ scaled to have norm $1$.
Choose $G$ large enough to have the matrix $Z=[J, \hat G_l,
\beta_q \hat G_q, \beta_h \hat H, N_G]$ with small condition
number (say less that $10^2$).

\item[{\bf C}] Find the minimal $l_1$ solution to the
underdetermined system $Y_n=Z\alpha$.

\item[{\bf D}] Choose a threshold $T_n$ and let $\alpha_{T_n}$ be
the coefficients in $\alpha$ larger than $T_n$. Let $\mathcal
I_n$, the estimated set of directed links of node $n$, be the
union of all node indexes that appear in terms of $Z$
corresponding to coefficients in $\alpha_{T_n}$.
\end{itemize}

Basically in step {\bf A} we use the sampled trajectories to
estimate the integrals in the representation on the right hand
side of equation (4). In step {\bf B} we scale the terms of the
representation, we attenuate quadratic and hyperbolic terms, and
we augment the model with scaled random terms. In step {\bf C} we
apply $l_1$ minimization. Finally in step {\bf D} we select the
largest coefficients in the representation and we estimate the set
of links that determine the dynamics of each node. The choice of
the threshold in step {\bf D} is very delicate and it is explored
in depth for the epidermal growth factor receptor signaling
network that we study in the next section. We stress again that by
no means we need to limit ourselves to linear, quadratic and
hyperbolic terms. general power function expansions or higher
degree polynomial terms are possible within the frame of this
method, since random terms and $l_1$ minimization keep the
reconstruction stable even for very underdetermined systems of
representation.

\section{Results}

\subsection{The epidermal growth factor receptor (EGFR) Network}

In this section we show the performance of the augmented sparse
reconstruction method {\bf A-D} on the epidermal growth factor
receptor (EGFR) protein network described in \cite{Araujoetal} and
explicitly shown in the appendix. We again emphasize that the
ability to dynamically measure and collect enough data from every
protein/node within the network is impossible with current
experimental methodologies. The EGFR network is one of the most
well-studied and clinically important signaling networks in
biology today and the ability of our method to reconstruct a model
of such fundamental network is very promising. The EGFR network
has been modelled in \cite{Araujoetal} as a system involving only
linear, quadratic and hyperbolic terms, so the general model in
equation (1) (and therefore in equation (4)) is ideally suited for
its analysis, however the right hand side of equations (1) and (4)
have a very large number of quadratic and hyperbolic terms for the
EGFR network, both due to the large number of variables involved,
and to the need of considering a sufficiently large sampling of
hyperbolic terms. Therefore already in this case we are faced with
the extreme difficulty of finding the few relevant terms for the
actual EGFR network. Despite this difficulty, augmented sparse
reconstruction is able to find a very significant fraction of the
links in the network. Moreover preliminary results show that the
method is robust with respect to changes in the size and type of
system of representation.

The sparsity of links for the EGFR system has some variation
between nodes; we have $11$ variables with less than $4$ distinct
terms (linear, quadratic or hyperbolic) in the expression for
their derivative, $9$ variables with less than $8$ terms and $1$
variable, $x_4$, with $19$ terms. This last variable is not sparse
and corresponds to the main `hub' of the EGFR network.

We assume that $100$ time series with different initial condition,
each of a length of $25$ points, are available for each variable
in the system. Only $500$ uniformly selected points among the
total $2500$ are used in the algorithm, so that we are effectively
working with a very small data set of points, even though we gain
some information from the missing points when estimating the
integrals in equation (4). The initial conditions for each
variable are chosen as uniformly distributed random numbers in the
interval $[0,\,40]$. In real systems the biologically significant
ranges of initial conditions vary among different variables. This
raises an interesting theoretical and practical question: which is
the minimal domain of initial conditions that allows the
reconstruction of the network? This question is particulary
relevant for networks that display simple dynamics, since each
short trajectory may not carry the full information on the
underlying network.

The length of the time series is chosen to be consistent with the
sampling rate that can be performed in practice, that is the
reason we take only $25$ uniformly spaced points in the time
interval $[0,\,27]$ along each trajectory. To put this number in
perspective, we note that for a time series with fast initial
decay like the one in Figure 1(a) this mean that we have only $2$
to $3$ points in the high varying region of the series, while for
a time series with slow decay as the one in Figure 1(b) we have a
larger proportion of points where the time series has not yet
relaxed to its steady state. As we already stressed, this
infrequent sampling is one of the reasons we had to move from the
differential representation of the network to the integral one
used in {\bf A-D}, as it may be problematic to estimate
derivatives in such infrequent sampling scenario. The way we add
noise to the trajectories is by taking the maximum $M$ of each
given time series and by adding uniform white noise in the
interval $[-m,\, m]$ where $m$ is equal to a fraction of $M$; this
seems to give levels of noise consistent with experimental
conditions. The characteristic shape of the noisy time series from
the EGFR network, is shown in Figure 1 (noise level $15\%$).

The sampling interval of the hyperbolic terms is $\bar c=10$ and
the total number of hyperbolic terms for each variable is $S=50$.
The total number of terms for the model, and therefore the total
number of parameters, is $1449$, far more than the $500$ data
points we use to find the links for each node. The number of
random vectors to be used in step {\bf B} is chosen as $G=2500$.
The attenuation for the quadratic and the hyperbolic terms is
chosen to be $\beta_q=\beta_h=0.5$.

%FIGURE 1
%
%
\begin{figure}
\includegraphics[angle= 0,width=1\textwidth]{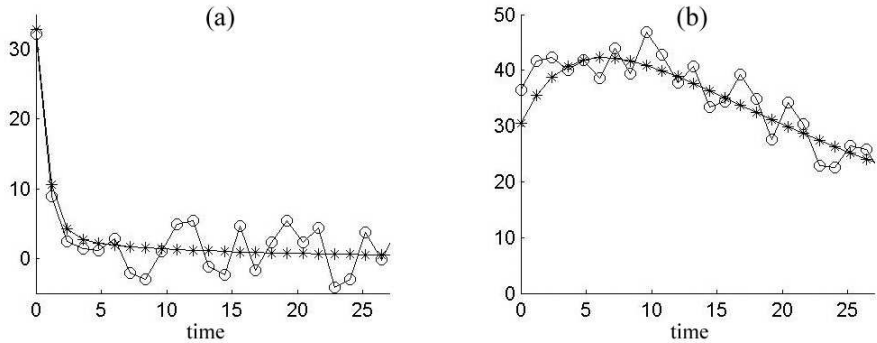}
\caption{In subplots (a) and (b) we show typical trajectories that
are observed in the  EGFR system, sampled uniformly $25$ times in
the time interval $[0,\,27]$. Starred curves are the actual
trajectories, circled curves are the trajectories with $15\%$
relative noise added. Plot (a) shows a trajectory of $x_9$ that
settles within few samples points to a base value, plot (b) a
trajectory of $x_6$ with a much slower decay.}
\end{figure}
In Figure 2 we show a typical example of the sparse representation
that can be obtained by applying {\bf A-D} to the infrequently
sampled, noisy trajectories of the EGFR network with the noise
level as in Figure 1. More specifically, we show the reconstructed
representation for $Y_2$, the vector of all integrals of
$\dot{x}_2$ defined in step {\bf A}, with respect to the integral
of all linear, quadratic, and hyperbolic terms, as defined in step
{\bf B}. We choose variable $x_2$ because it has very few terms in
its actual representation of the derivative, namely
$\dot{x}_2=-0.06x_2+0.2x_3+0.003x_1x_{23}-0.02x_2^2$, having a
sparsity for which the algorithm works often at its best. We plot
the norm of the coefficients of: the linear terms in Figure 2(a),
from $G_1$ to $G_{23}$; the quadratic terms in Figure 2(b),
ordered as $G_{1,1}$,..., $G_{1,23}$, $G_{2,2}$,..., $G_{22,23}$;
the hyperbolic terms in Figure 2(c), ordered as $H_{1,1}$,...,
$H_{1,10}$,..., $H_{23,1}$,..., $H_{23,10}$; and the random terms
in Figure 2(d).
%
%
%FIGURE 2
%
\begin{figure}
\includegraphics[angle= 0,width=1\textwidth]{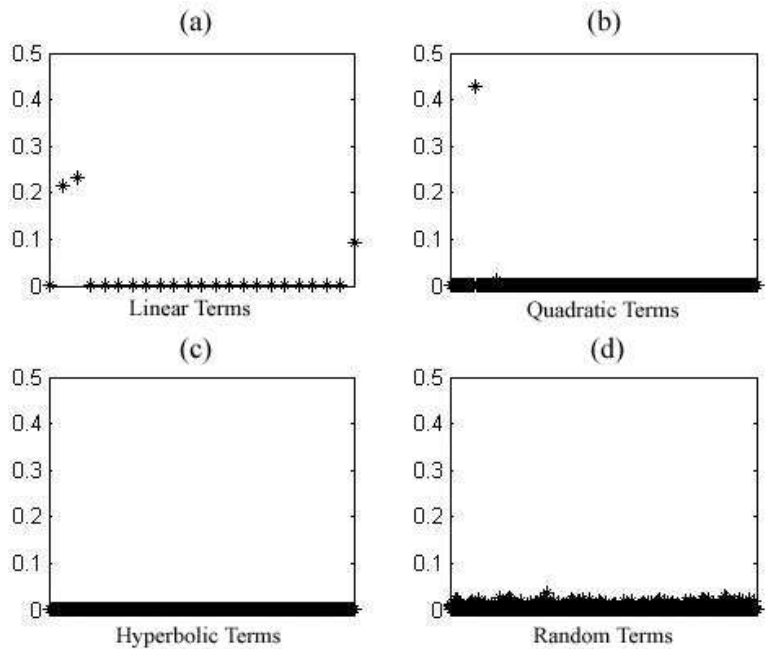}
\caption{From top left, we plot the norm of the coefficients of
$Y_2$ (as defined in step {\bf A}) for: (a) the linear terms $\hat
G_l$; (b) the quadratic terms $\hat G_{q}$; (c) the hyperbolic
terms $\hat H$; and (d) the random terms.}
\end{figure}
The $3$ largest coefficients across all terms correspond exactly
to three of the terms in the representation of $\dot{x}_2$, namely
$x_2$, $x_3$ and $x_1x_{23}$, we are missing instead the $x_2^2$
term. The forth largest coefficient in the reconstructed
representation corresponds to the $x_{23}$ term, so it repeats to
some extent the information on the network linkage given by the
$x_1x_{23}$ term. If we apply the reconstruction algorithm to this
node with $20$ different realizations of $15\%$ relative noise,
the $x_2$ term appears as dominant $18$ times, the $x_3$ term $15$
times, the $x_1x_{23}$ term $20$ times and the $x^2$ term $3$
times. Note that the dominance of a term is not only due to the
size of its coefficient, but to the intensity of the corresponding
signal as well, for example $x_1x_{23}$ has coefficient $0.003$ in
the equation for $\dot x_2$ and yet it is recovered more often
than $x^2$ that has larger coefficient $0.02$. We believe this has
to do with the specific norm scaling that is selected in step {\bf
B} of the algorithm. Different choices of norm scaling may be
useful to improve further the performance of the method.

The example of the reconstruction of the representation of $\dot
x_2$ is typical: some terms not only may be missing, but they can
be partially wrong, for example a term $x_i$ may appear in the
representation as $x_i^2$, or a term $x_ix_j$ may be replaced by a
term $x_ix_k$ that gives similar shapes for the given initial
conditions. Note that these two possibilities do carry some
significant information on the geometry of the network, even
though the specific terms are incorrect.

We can  compare at this point the effectiveness of our $l_1$
method in identifying the relevant links with respect to simpler
techniques such as correlation. In Figure 3 we show, for
comparison, the correlation $Y_2$ with: the linear terms in Figure
3(a), from $G_1$ to $G_{23}$; the quadratic terms in Figure 3(b),
ordered as $G_{1,1}$,..., $G_{1,23}$, $G_{2,2}$,..., $G_{22,23}$;
the hyperbolic terms in Figure 3(c), ordered as $H_{1,1}$,...,
$H_{1,10}$,..., $H_{23,1}$,..., $H_{23,10}$. The most negatively
correlated linear term corresponds to $x_2$, the most negatively
correlated quadratic term to $x_2^2$, and the cluster of most
negatively correlated hyperbolic terms correspond to $x_2$ as
well. Note however that many terms show similar level of large
negative correlation, especially among  the quadratic terms, so
that it is difficult to set a threshold on the norm of the
correlation coefficients that would, for example, identify {\it
only} $x_1x_{23}$ as another relevant term. The key point is that
the considerable sparsity of the reconstructions computed by our
method allows for an accurate distinction of false links and true
links. We do not explore this issue further in this paper, but see
our forthcoming paper \footnote{D. Napoletani, T. Sauer,
Reconstructing the Topology of Sparsely-Connected Dynamical
Systems, submitted.} for a comparison of the augmented sparse
reconstruction with $l_2$ regression on a special class of ODE
networks.
%FIGURE 3
%
\begin{figure}
\includegraphics[angle= 0,width=1\textwidth]{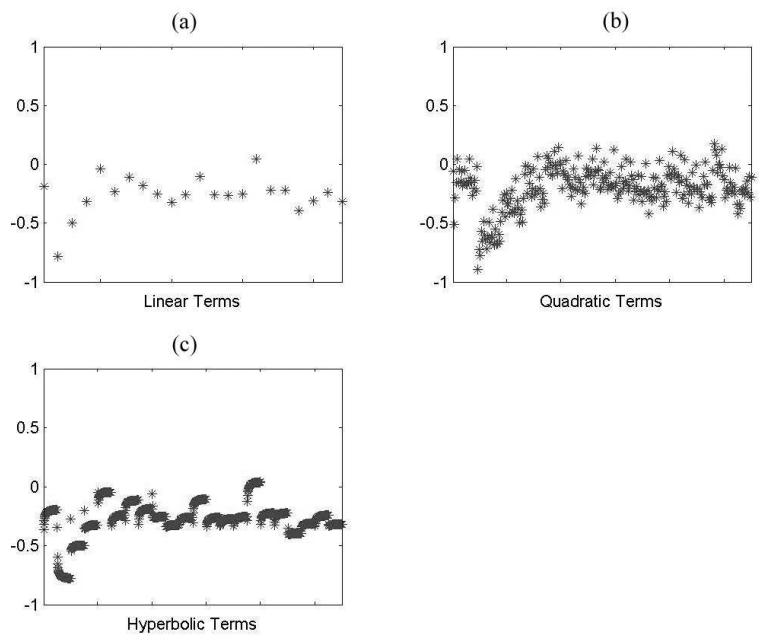}
\caption{From top left, we plot the correlation coefficients of
$Y_2$ (as defined in step {\bf A}) with: (a) the linear terms $\hat
G_l$; (b) the quadratic terms $\hat G_{q}$; (c) the hyperbolic
terms $\hat H$.}
\end{figure}

To evaluate globally the quality of the reconstruction results for
different levels of noise we use the ratio of computed true links
with respect to the total number of true links (true positives
rate) and the ratio of computed false links with respect to the
total number of false links (false positives rate). An important
question when assessing the quality of reconstruction is the
proper estimation of the thresholds $T_n$ used in {\bf D}. In
general we expect these thresholds to vary according to the noise
level in the time series, but even the sampling rate will affect
our degree of confidence in the computed links so we must find an
automatic way to estimate the threshold from the data. Note
moreover that the threshold must be represented in terms of the
coefficients used to represent each node. To this extent we define
a threshold, for a given system, as a constant multiple of the
standard deviation of the non-zero coefficients of the non-random
terms of {\it each node}, what we may call the deterministic
coefficients of the representation (in practice we neglect any
coefficient with norm smaller than $10^{-10}$). Formally we define
$T_i=K\sigma_i$, $i=1,...,N$, where $K$ is some fixed constant
determined for the whole system, while $\sigma_i$ is the standard
deviation of the absolute value of the deterministic non-zero
coefficients of the representation of node $i$. This flexible
definition of the threshold ensures that: a) the threshold level
is relative to the norm of the coefficients of each node; b) the
threshold is larger if there are many sizeable non-zero
coefficients in the representation of a specific node. The main
advantage of a uniform definition of threshold across all
variables is that we need the proper estimation of a single
threshold $K$, and we have the whole reconstruction data available
to do that. If the network has very distinct behavior for
different subsets of nodes, it may not be possible to use a single
multiplier and we must resort to thresholds estimated for each
node separately.

Before suggesting a specific way to find the uniform threshold $K$
from the computed representations, let us see what we would get
with an `ideal' choice of it. Suppose that for each noise level in
the time series, we select $K$ so that the false positives rate
stays below $0.1$. We perform such analysis for $20$ realizations
for each level of relative noise in the trajectories from $0\%$ to
$25\%$. In Figure 4(a) we can see the result of such choice of
thresholds: the average true positives rate (starred curve) is
high (around $0.65$) even for realistic trajectories' noise of the
order of $20\%$. For all $20$ realizations the computed true
positive rates differed from the average by at most $0.054$ units,
with standard deviation, at each noise level, of at most 0.026.
The computed average values of $K$ are: $K_{0\%}=0.032$,
$K_{5\%}=0.046$, $K_{10\%}= 0.067$, $K_{15\%}=0.099$,
$K_{20\%}=0.120$, $K_{25\%}=0.137$. The noisier the time series,
the higher the value of $K$ needed to keep the false positives
rate small.
%FIGURE 4
%
\begin{figure}
\includegraphics[angle= 0,width=0.8\textwidth]{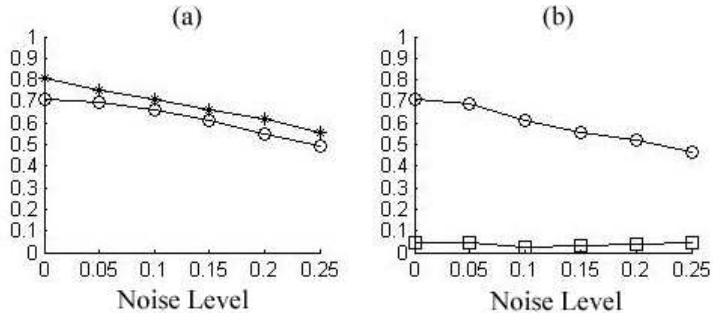}
\caption{In plot (a) we have the average true positives rates for
relative noise in the trajectories from $0\%$ to $25\%$ when the
value of the threshold multiplier $K$ is artificially set to keep
the false positive rate at: $0.1$ (starred curve) and $0.05$
(circled curve). In plot (b) we have the average true positive
rates (starred curve) and average false positive rates (squared
curve) for relative noise in the trajectories from $0\%$ to $25\%$
where the value of the threshold multiplier $K$ is found for each
noise level by using the heuristic {\bf E}.}
\end{figure}
%rateTP 0.1=
%
%    0.8072
%    0.7532
%    0.7095
%    0.6604
%    0.6176
%    0.5586
Figure 4(a) shows also the average true positives rates (circled
curve) when the false positives rate is kept at $0.05$. In this
case, for all $20$ realizations the computed true positive rates
differed from the average by at most $0.071$ units, with standard
deviation, at each noise level, of at most 0.03. The computed
average values of $K$ are: $K_{0\%}=0.104$, $K_{5\%}=0.129$,
$K_{10\%}= 0.185$, $K_{15\%}=0.253$, $K_{20\%}=0.315$,
$K_{25\%}=0.369$.
%rateTP 0.05=
%
%    0.7122
%    0.6982
%    0.6581
%    0.6135
%    0.5514
%    0.4950
Even for noiseless trajectories we still do not find all true
links, this is due in part to the infrequent sampling of the
trajectories that hides subtle interactions among the nodes. Note
that the rates we display are obtained excluding from the average
the reconstruction of variable $4$ that does not have sparse
representation and which is a `hub', so likely to be better knwon
experimentally. We decided to exclude that specific protein in our
estimates of the errors because only very few proteins are
believed to perform the role of hub of a network in signaling
pathways, therefore we believe that the error rates computed above
are a better indicator of the errors we would find in computing
the representation of a generic protein in a large network.
Moreover the representation of $x_4$ is so different from the
others that the use of the same threshold multiplier for it as for
all other variables does not seem appropriate. Including variable
$4$ in computing the errors, without any modification in the
choice of threshold multiplier, would lead true positive rates to
slightly worsen.

It is possible in principle to have cases when much finer sampling
of the trajectories is experimentally available. To simulate this
scenario, suppose we sample the trajectories of the EGFR network
uniformly $100$ times in the time interval $[0,\,27]$, then the
error rates improve significantly. If we keep false positive rate
to $0.1$, then average true positive rates are about $0.88$ in the
absence of noise, this is a improvement of almost $0.08$ with
respect to the same time series sampled uniformly only $25$ times.
In Figure 5(a) we show average true positive rates for noise level
from $0\%$ to $25\%$ in this fine sampling scenario. We use $20$
realizations for each noise level in computing the averages

%figure 5
%
\begin{figure}
\includegraphics[angle=0,width=1\textwidth]{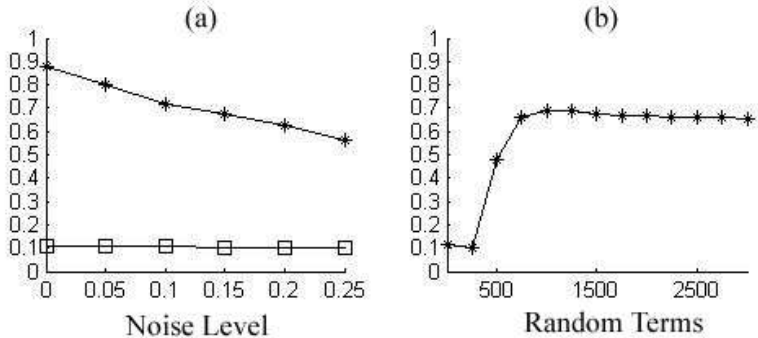}
\caption{In plot (a) we have the true positives rates (starred
curve) for relative noise in the trajectories from $0\%$ to $25\%$
when the value of the threshold multiplier $K$ is artificially set
to keep the false positive rate (squared curve) at $0.1$. Network
trajectories are sampled $100$ times in the interval $[0,\,27]$ to
generate this plot. In plot (b) we have the true positive rates
(starred curve) as a function of the number of random terms added
to the model. Network trajectories used to generate this plot are
sampled $25$ times in the interval $[0,\,27]$ and noise level is
kept at $15\%$. The value of the threshold multiplier $K$ is
artificially set to keep the false positive rate at $0.1$.}
\end{figure}
Figure 5(b) shows that a large number of random terms is important
for the proper functioning of algorithm {\bf A-D}. Namely we show
that the average true positive rates improve when we increase the
number $G$ of random terms from $0$ to $3000$ in intervals of
$250$. Trajectories are sampled infrequently ($25$ times in the
interval $[0,\,27]$), false positive rates are kept at $0.1$ and
relative noise level is fixed at $15\%$. The true positives rate
is very low, just $0.13$, when there are no random terms added.
Addition of  $1000$ or more terms gives high true positive rates,
about $0.68$, that are comparable for several values of the number
of terms $G$. These results show also the robustness of the
algorithm with respect to the choice of $G$.

\subsection{Choice of Threshold}
We now approach the problem of finding a suitable value of $K$
from the reconstruction data generated by the algorithm itself at
the end of step {\bf C}. Denote by $S(K)$ the total number of
selected links that are found in step {\bf D} of the augmented
sparse reconstruction algorithm by using thresholds
$T_n=K\sigma_n$. We can split $S(K)$ as $S(K)=S_t(K)+S_f(K)$ where
$S_t(K)$ denotes the number of true computed links and $S_f(K)$
the number of false computed links. Since for each node we have
only a small number of true links by assumption, and their
corresponding coefficients in the representation are, in general,
very large, we can conjecture that, as we let $K$ increase
continuously from $0$ to $\infty$, $S_t(K)$ will decrease very
slowly at the beginning. Since $S_t(K)$ assumes only integer
values, this slow decay will appear as infrequent small jumps,
this means that the (discontinuous) derivative $dS(K)$ of $S(K)$
will be dominated by the derivative $dS_f(K)$ of $S_f(K)$ for
small values of $K$ and by $dS_t(K)$, the derivative of $S_t(K)$,
for larger values of $K$, therefore we can infer some of the
properties of $S_f(K)$ which is not known, from those of $S(K)$,
which is a computable function.

In Figure 6(a) to 6(c) we show approximations to $dS_t(K)$,
$dS_f(K)$ and $dS(K)$ for a specific reconstruction with relative
noise in the time series of the order of $10\%$. We choose a fine
uniform sampling $U=0.001$ of $K$, up to $K=3.5$, so that $dS(K)$
never goes below $-2$. To have the ideal case in which $dS(K)\geq
-1$ for all $K$ seems to require excessively fine sampling rate.
Note that $S(K)$ is identically zero for $K>3.26$, we can also
immediately see the similarity of $dS(K)$ and $dS_f(K)$ in the
frequency of negative jumps for small values of $K$. The frequency
of jumps of $dS(K)$ greatly decreases around $K=0.30$, to see this
transition point more clearly, let $K_1,...,K_M$ be the values of
$K$, ordered from smallest to largest, for which $dS(K)\neq 0$ and
define a function $J(i)=K_i-K_{i-1}$, $i=2,...,M$ that computes
the width of negative jumps. We plot $J$ in Figure 6(d) and we can
see that for $i\approx 57$ we suddenly have much wider intervals
between jumps, this value of $i$ corresponds to $K_{57} \approx
0.29$. We argue that a suitable value $K_f$ of the multiplier $K$
is the one for which $J(i)$ has very different local averages for
$i<f$ and $i>f$. To find such $K_f$ we can use the following rule:
%
%%0.291 for 57
%
%FIGURE 6
%
\begin{figure}
\includegraphics[angle= 0,width=1\textwidth]{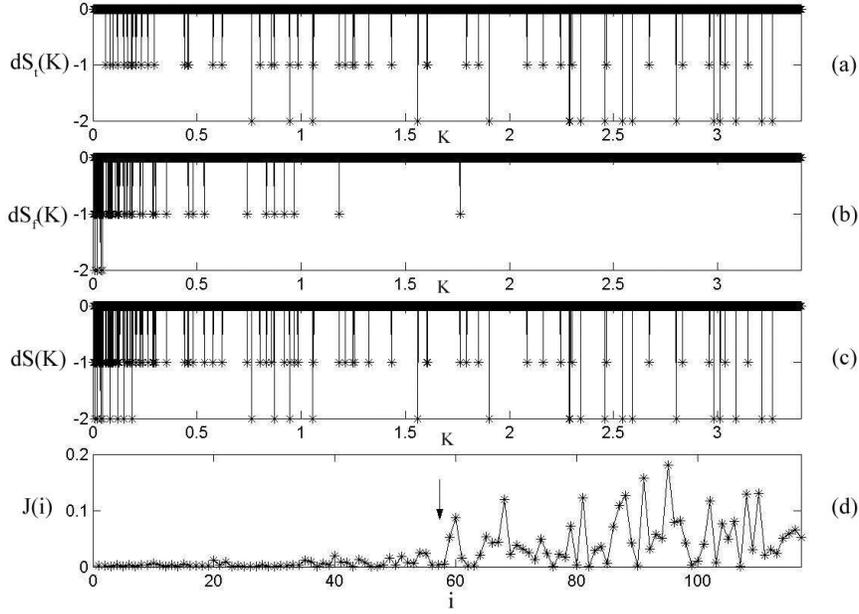}
\caption{Plots of:(a) $dS_t(K)$; (b) $dS_f(K)$; (c) $dS(K)$ for
$10\%$ relative noise in the trajectories. In plot (d) we show
$J(i)$, the distance between the $i-1$-th and the $i$-th negative
jumps in $dS(K)$, for $10\%$ relative noise in the trajectories.
$K$ is sampled uniformly with sampling interval of size $U=0.001$.
The arrow in plot (d) points to the index values, around $i=57$,
for which we have a large change of mean frequency of jumps.}
\end{figure}

\begin{itemize}
\item[{\bf E}] Set an integer I, let $V(i)=\frac{\bar
J_{i-I}}{\bar J_{i+I}}$, $i=I,...,M-I$, where we denote by $\bar
J_{i-I}$ the mean of $J$ for values between $i-I$ and $i$ and by
$\bar J_{i+I}$ the mean of $J$ for values between $i$ and $i+I$.
Denote by $f$ the index for which $V(i)$ is minimum, and by $K_f$
the corresponding threshold multiplier.
\end{itemize}
Rule {\bf E} uses the ratio of the local mean of the length of
intervals between jumps before and after the $i$-th jump to select
the jump for which the relative increase is maximum. If we use
$I=20$ in {\bf E}, we find from the output data of step {\bf C} of
the algorithm the following threshold multipliers:
$K_{0\%}=0.114$, $K_{5\%}= 0.164$, $K_{10\%}=0.328$,
$K_{15\%}=0.415$, $K_{20\%}=0.428$, $K_{25\%}=0.486$. In Figure
4(b) we plot the average true positive rates and average false
positive rates computed by using in step {\bf D} of the algorithm
by using these estimated values of $K$. We use $20$ realizations
for each noise level in computing the averages, for all $20$
realizations the computed true positive rates differ from the
averages by at most $0.11$ units, with standard deviation, at each
noise level, of at most 0.058. The false positives rate is below
$0.11$ for all realizations and all noise levels. With $10\%$
relative noise in the trajectories, we have a significant average
true positives rate of about $0.61$ with average false positives
rate of only $0.025$. There is a greater variability in true
positives and false positives rates in Figure4(b) with respect to
those computed in Figure4(a), this fact points out the need of a
more sophisticated analysis of the change of mean frequency. For
example, a wavelet maxima analysis (\cite{Mallat}, chapter 6) of
$J(i)$ can be used for a more robust evaluation of the thresholds.
It would be of great interest to deduce theoretically the value of
$K_f$, under suitable conditions on the class of network models.
%%%
%rateFP =
%
%    0.0485
%    0.0477
%    0.0253
%    0.0298
%    0.0398
%    0.0446
%
%rateTP =
%
%    0.7108
%    0.6874
%    0.6104
%    0.5532
%    0.5189
%    0.4671
%%
%
%
%

\section{Discussion}

There are still many open questions whose answers will shape the
way augmented sparse reconstruction methods are applied to network
reconstruction. What is the limiting node sparsity that still
allows the network itself to be recovered with this method? how
are the error rates affected if only a subset of the variables is
available? how does the network we compute on this subset of
variables relates to the full network? if we have a node that is
unrelated to the chosen subset of the network, it tends to have a
greater portion of its norm accounted for by the random terms, and
this ca be used to decide whether it is well connected to the
other variables, but further research in this direction is needed.

In some cases the `skeleton' of the network may be available, for
example for proteins networks we may know roughly how the system
is connected for healthy cells. Can we use this additional
information to detect, with this method, whether patients with
cancer develop additional strong links among nodes? Preliminary
evidence suggests that small new links that do not make the system
unstable are often detectable, but it would be interesting to use
the available information on the skeleton of the network directly
in the algorithm.

Even when previous information on the network is not available, it
is very important for clinical applications to determine whether a
specific protein has very distinct representations for cancerous
cells and for healthy ones. If this is the case, then the
reconstruction algorithm can predict changes in the signaling
pathways that are likely due to the cancer itself.

One possible extension of our method is to perform the estimation
of links only on local subsets of trajectories. This local
application of the algorithm may highlight different links that
could be dominant for different sets of initial conditions, in
\footnote{D. Napoletani, T. Sauer, Reconstructing the Topology of
Sparsely-Connected Dynamical Systems, submitted.} we show that
this strategy is indeed feasible. Note that the augmented sparse
reconstruction scheme has an edge over simple $l_2$ regression
especially when there is a very limited set of initial conditions.
Even in the case in which it is possible to span the entire phase
space of the network, there is a limit to the density of the
initial conditions that can be taken and therefore a local
application of this method will be beneficial (because there are
only a few local trajectories). By putting together the
information on links that arise in different regions of the phase
space it may be possible to find very tenuous links that would
otherwise be undetectable in a global analysis. A clear advantage
of a local version of the method is its generality, since simple,
low degree polynomial models can always be used.

The augmented sparse reconstruction described in this paper is
able to identify relevant links among nodes in very large systems
of representation and with very noisy conditions, so we expect the
algorithm to scale well to the use of cubic or higher degree terms
in the representation and even to the use of general power
functions, possibly with non integer and negative exponents. The
use of attenuation of blocks of terms in the representation will
turn out to be even more important when the size of the dictionary
of terms is increased.

We stated in the introduction that a promising approach to
biological networks is to deemphasize exact modeling, in favor of
a robust identification of classes of suitable models. If we take
one step forward in this direction, then the techniques of network
control, and the very notion of global stability of a network,
must be changed in such a way that they are valid for entire
classes of indistinguishable systems \cite{Smith} that produce
trajectories that are qualitatively similar. In this perspective,
The reconstruction algorithm described in this paper could be used
as an intermediate step of data-driven control schemes based on
particle filter techniques, by providing an indistinguishable
model that locally behaves as the real one. This potential
application of the augmented sparse reconstruction method would be
an interesting step in the direction of real time, personalized
therapies that require an online estimation and control of
specific pathways in the cell networks of individual patients,
\cite{LP1}, \cite{LP2}.

\section*{Appendix: the EGFR network}
\begin{align*}
&\dot{x}_1=0.06x_2-0.003x_1x_{23}\\
&\dot{x}_2=-0.06x_2+0.2x_3+0.003x_1x_{23}-0.02x_2^2\\
&\dot{x}_3=-1.1x_3+0.01x_4+0.01x_2^2\\
&\dot{x}_4=x_3-0.01x_4+0.2x_5+0.3x_6+0.05x_7\\
&+0.03x_8+0.6x_9+0.3x_{10}+0.3x_{11}+0.12x_{12}\\
&-0.0045x_4x_{13}-0.0009x_4x_{14}-0.0009x_4x_{15}\\
&-0.06x_4x_{16}-0.006x_4x_{17}-0.003x_4x_{19}-0.09x_4x_{20}\\
&-0.00024x_4x_{22}-\frac{450x_4}{50+x_4}\\
&\dot{x}_5=-1.2x_5+0.05x_6+0.06x_4x_{16}\\
&\dot{x}_6=x_5-0.35x_6+0.006x_4x_{17}\\
&\dot{x}_7=-0.05x_7+0.06x_8-0.01x_7x_{21}+0.003x_4x_{19}\\
&\dot{x}_8=-0.09x_8+0.0045x_4x_{13}+0.01x_7x_{21}\\
&\dot{x}_9=-6.6x_9+0.06x_{10}+0.09x_4x_{20}\\
&\dot{x}_{10}=6x_9-0.07x_{10}+0.0009x_4x_{14}\\
&\dot{x}_{11}=-0.4x_{11}+0.0214x_{12}+0.0009x_4x_{15}-0.01x_{11}x_{21}\\
&+0.003x_{10}x_{19}\\
&\dot{x}_{12}=-0.1843x_{12}+0.00024x_4x_{22}+0.009x_{10}x_{13}+0.01x_{11}x_{21}\\
&\dot{x}_{13}=0.03x_8+0.0429x_{12}-0.0015x_{13}+0.1x_{22}\\
&-0.0045x_4x_{13}-0.009x_{10}x_{13}-0.021x_{13}x_{14}\\
&+0.0001x_{19}x_{21}\\
%\end{align*}
%\begin{align*}
&\dot{x}_{14}=0.3x_{10}+0.1x_{15}+0.1x_{22}-0.0009x_4x_{14}\\
&-0.021x_{13}x_{14}-0.003x_{14}x_{19}-\frac{1.7x_{14}}{340+x_{14}}\\
&\dot{x}_{15}=0.3x_{11}-0.1x_{15}+0.064x_{22}-0.0009x_4x_{15}\\
&+0.003x_{14}x_{19}+0.03x_{15}x_{21}\\
&\dot{x}_{16}=0.2x_{5}-0.06x_4x_{16}+\frac{x_{17}}{100+x_{17}}\\
&\dot{x}_{17}=x_{6}+x_{17}+x_{18}+x_4x_{17}+\frac{x_{17}}{1+x_{17}}\\
&\dot{x}_{18}=x_{17}-0.03x_{18}\\
&\dot{x}_{19}=0.05x_{7}+0.1x_{11}+0.0015x_{13}+0.1x_{15}\\
&-0.003x_4x_{19}-0.003x_{10}x_{19}-0.003x_{14}x_{19}\\
&-0.0001x_{19}x_{21}\\
&\dot{x}_{20}=0.6x_{9}-0.09x_4x_{20}+\frac{1.7x_{14}}{340+x_{14}}\\
&\dot{x}_{21}=0.06x_{8}+0.0214x_{12}+0.0015x_{13}+0.064x_{22}\\
&-0.01x_7x_{21}-0.01x_{11}x_{21}-0.03x_{15}x_{21}\\
&-0.0001x_{19}x_{21}\\
&\dot{x}_{22}=0.12x_{12}-0.064x_{22}-0.00024x_4x_{22}+0.021x_{13}x_{14}\\
&+0.03x_{15}x_{21}\\
&\dot{x}_{23}=0
\end{align*}

\end{document}